\RequirePackage{fix-cm}
\documentclass[smallextended]{svjour3}       
\smartqed  
\usepackage{graphicx}
\usepackage{epstopdf}
\usepackage{geometry}
\usepackage{graphicx,amssymb,amstext,amsmath}
\usepackage[justification=centering]{caption}

\usepackage{epsfig}
\usepackage{subfigure}
\usepackage{amsmath}
\usepackage{algorithmic}
\usepackage{graphics}
\usepackage{amssymb}
\usepackage{alltt}
\usepackage{float}
\usepackage{array}
\usepackage{xspace}
\usepackage{footmisc}
\usepackage{comment}
\usepackage{algorithm}
\usepackage{algorithmic}
\usepackage[misc]{ifsym}

\DeclareMathOperator*{\argmax}{arg\,max}

\begin{document}

\title{Efficient Interactive Search for Geo-tagged Multimedia Data
}
\subtitle{}


\author{Jun Long         \and
        Lei Zhu          \and
        Chengyuan Zhang  \and
        Zhan Yang      \and
        Yunwu Lin      \and
        Ruipeng Chen       \and
}


\institute{Jun Long \at
              \email{jlong@csu.edu.cn}           
           \and
           Lei Zhu \at
              \email{leizhu@csu.edu.cn}
           \and
           \Letter Chengyuan Zhang \at
              \email{cyzhang@csu.edu.cn}
           \and
           Zhan Yang \at
              \email{zyang22@csu.edu.cn}
           \and
           Yunwu Lin \at
              \email{lywcsu@csu.edu.cn}
           \and
           Ruipeng Chen \at
              \email{rpchen@csu.edu.cn}
           \\
           \\
           School of Information Science and Engineering, Central South University, PR China \\
           Big Data and Knowledge Engineering Institute, Central South University, PR China
}

\date{Received: date / Accepted: date}

\maketitle

\begin{abstract}
Due to the advances in mobile computing and multimedia techniques, there are vast amount of multimedia data with geographical information collected in multifarious applications. In this paper, we propose a novel type of image search named interactive geo-tagged image search which aims to find out a set of images based on geographical proximity and similarity of visual content, as well as the preference of users. Existing approaches for spatial keyword query and geo-image query cannot address this problem effectively since they do not consider these three type of information together for query. In order to solve this challenge efficiently, we propose the definition of interactive top-$k$ geo-tagged image query and then present a framework including candidate search stage , interaction stage and termination stage. To enhance the searching efficiency in a large-scale database, we propose the candidate search algorithm named GI-SUPER Search based on a new notion called superior relationship and GIR-Tree, a novel index structure. Furthermore, two candidate selection methods are proposed for learning the preferences of the user during the interaction. At last, the termination procedure and estimation procedure are introduced in brief. Experimental evaluation on real multimedia dataset demonstrates that our solution has a really high performance.
\end{abstract}

\section{Introduction}
\label{Intro}

With the rapid development of mobile Internet and social multimedia applications, huge amount of multimedia data~\cite{DBLP:conf/mm/WangLWZZ14} such as text, image, audio and video have been generated and shared everyday. For instance, Facebook\footnote{https://facebook.com/} reports 350 million photos uploaded daily as of November 2013. More than 400 million daily tweets containing texts and images have been generated by 140 million Twitter\footnote{http://www.twitter.com/} active users. Flickr\footnote{http://www.flickr.com/} had a total of 87 million registered members and more than 3.5 million new images uploaded daily in March 2013. Large-scale of multimedia data stored in massive databases lead to multifarious innovative Internet services.

Just as an English idiom said, a picture is worth a thousand words, that means a image contains much more information than a sentence or a few words. Keyword-based image retrieval methods have the limitation that they have to depend on manual annotation. Obviously, it is impractical that annotating all image data in a large-scale database manually. Besides, we cannot use a few of keywords to describe most of photos comprehensively. Unlike this traditional technique, content-based image retrieval (CBIR for short) has extensive applications, which use inherent visual content of images for searching and query. Over the last decade, lots of researchers have been attracted by CBIR techniques in the area of multimedia information retrieval~\cite{DBLP:conf/cikm/WangLZ13,DBLP:journals/tnn/WangZWLZ17} and many CBIR systems like K-DIME~\cite{DBLP:journals/ieeemm/Bianchi-Berthouze03},IRMFRCAMF~\cite{DBLP:journals/remotesensing/LiZTZ16}, gMRBIR~\cite{DBLP:journals/prl/ChenWcYM16} have been developed for constructing multimedia intelligent systems.

Modern mobile computing devices like smartphones and tablets are equipped with high-performance image processor and high-resolution digital camera which are supported by advanced camera applications. This makes it possible that the users can take high-quality photos and share them to friends anytime and anywhere. In addition, as the locating techniques are widely applied in smart devices, such as GPS modules, wireless communication modules, gyroscope sensors, compasses and etc., the mobile systems and applications can easily record geographical position information (latitude and longitude) so long as the devices are power-on. Combining both two techniques aforementioned, therefore, a new application scene is generated: people use their smartphones to take photos with current geographical information and share them in Internet platforms or social multimedia services such as Flickr, Instgram\footnote{http://instagram.com/}, Twitter, WeChat\footnote{https://weixin.qq.com/}, etc., and others can search and browse these artworks by visual content and geo-location information.

Previous geo-tagged image search techniques consider geographical information and visual content of images together~\cite{DBLP:conf/dasfaa/ZhaoKSXWC15}, but they neglect user's preference which is a particularly significant information in a query processing. Apparently, preference is the substantive representation of user's willingness when she conduct a query. Below is a motivating example in which both the visual content similarity, the geo-location of the results and the preferences of the user are considered.

\begin{figure}[thb]
\newskip\subfigtoppskip \subfigtopskip = -0.1cm
\centering
\includegraphics[width=0.8\linewidth]{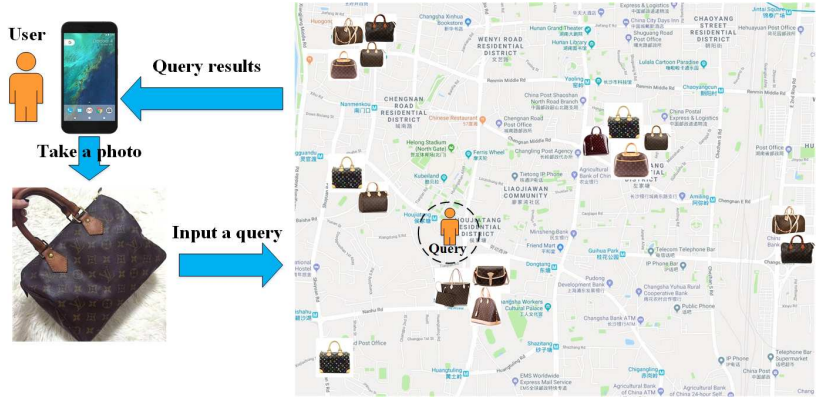}
\vspace{-1mm}
\caption{\small  An example of geo-tagged image query }
\label{fig:fig1}
\end{figure}

\begin{example}
\label{ex:example1}
As illustrated in Fig.~\ref{fig:fig1}, an user take a fancy to a favourite handbag and she would like to buy. However, she cannot describe it by language accurately and she do not know which nearby shop has the similar style of handbags. In such case, she can take a photo of the desirable handbag by her smartphone and input a geo-tagged image query to a image retrieval system. Then the system returns a set of relevant pictures with geographical location as the query result that can tell the user which shops have this kind of handbag and are close to her location. The user do not need to tell the system straightforwardly her preference about handbag such as color, size, handle style, shoulder strap style, etc., she just only to choose one favourite from the results and submit it again. The system will learn the user's preference according to the feedback and return another result set which is closer to her fondness. After a few times interaction, the system will find out desirable results for the user.
\end{example}

This work aims to implement the challenging application described in example~\ref{ex:example1}, namely, searching for a set of geo-tagged image from the database after several rounds of interaction according to the user's feedback. Based on the spatial keyword query techniques, we propose a novel query named interactive top-$k$ geo-tagged image query and design a new rank function named preference function which consider the preferences of users about both visual content and geographical proximity. In this paper we apply bag-of-visual-words model for image representation and a novel similarity measure is proposed based on the weight set similarity measure, which considers user's preference presented as a vector. In order to improve the efficiency of candidate searching in a large-scale image database, we propose a novel index structure named GIR-Tree which is based on bag-of-visual-words model and IR$^2$-Tree. Moreover, candidate search algorithm is proposed to address the challenge. To provide high performance of interactive query, we introduce a framework of interaction process to refine the candidate set in a iterative way and finally return a set of constraints from user's feedback. In the termination stage, we are enlightened by the support vector machine (SVM) techniques and design the preference estimation method which can generate a estimator of user's preference according to the constraints. At last the system can find out the more desirable results by preference estimator.

\noindent\textbf{Contributions.}
Our main contributions can be summarized as follows:
\begin{itemize}
\item To the best of our knowledge, we are the first to study interactive geo-tagged image search. We propose the definition of interactive geo-tagged image search and a novel score function based on user preference, and then design a framework including three stages for this problem.
\item For the candidates search before interaction, we define the superiority relationship between geo-tagged image objects at first. Then we propose a novel index structure named GIR-Tree based on IR$^2$-Tree and a efficient geo-tagged image superior search algorithm named GI-SUPER search.
\item For the strategy of interaction, we present the interaction algorithm and introduce two candidate selection methods which are the key procedure in this stage. Moreover, we propose the candidate filtering method.
\item We have conducted extensive experiments on real geo-tagged image dataset. Experimental results demonstrate that our proposed approach achieve high performance.
\end{itemize}

\noindent\textbf{Roadmap.} The remainder of this paper is organized as follows: We review the related work in Section 2. Section 3 introduces the basic techniques of image representation ,which are used in this paper and then formally defines the problem of interactive geo-tagged image search as well as relevant notions. In section 4, we propose a novel index structure named GIR-Tree index and then introduce the candidate search algorithm. Section 5 and 6 respectively present the interactive query strategy and termination procedure. Our experimental results are presented in Section 7, and finally we draw our conclusion of this paper in Section 8.

\section{Related work}
\label{related}
In this section, we present an overview of existing researches of image retrieval and ranking, spatial keyword query and interactive search, which are related to our study. To the best of our knowledge, there is no existing work on the problem of interactive geo-tagged image search.

\subsection{Image Retrieval and Interactive Search}
Image retrieval is one of the key techniques in the field of multimedia systems and retrieval~\cite{DBLP:conf/pakdd/WangCLZ13,DBLP:conf/cvpr/KimSX14,DBLP:journals/pami/YangNXLZP12,DBLP:conf/cvpr/DouzeRS11,DBLP:journals/corr/abs-1708-02288,DBLP:journals/ivc/WuW17}. It aims to find out a set of images~\cite{DBLP:journals/cviu/WuWGHL18} from a multimedia database, which are relevant to a query~\cite{DBLP:journals/mta/WuHZSW15,DBLP:journals/prl/ChenWcYM16,DBLP:conf/cvpr/DengBL11}.

Most existing researches on this field mainly focus on increasing accuracy or efficiency of the Retrieval results~\cite{DBLP:journals/tip/WangLWZZH15,DBLP:conf/cvpr/HuangLZM10,DBLP:journals/tnn/WangZWLZ17,TC2018}. For the problem that low-quality instances or even mistaken learning instances are often selected, Li et al.~\cite{DBLP:journals/mta/LiBZMWWW17} proposed a learning instance optimization method to select the optimized learning instances, which improved image retrieval quality. In order to increase the accuracy of landmark image retrieval~\cite{DBLP:journals/tip/WangLWZ17}, Zhu et al.~\cite{DBLP:journals/tcyb/ZhuSJZX15} proposed multimodal hypergraph (MMHG) to characterize the complex associations between landmark images. Based on MMHG, they developed a novel content-based visual landmark search system to facilitate effective image search. Wang et al.~\cite{DBLP:conf/mm/WangLWZ15} designed another approach to solve this problem. They analyzed a phenomena that the same landmarks provided by different users may convey different geometry information depending on the viewpoints and/or angles, and then proposed a novel framework named multi-query expansions to retrieve semantically robust landmarks by two steps. Driven by the problem that existing methods may not fully discover the geometric structure of the image set, Wang et al.~\cite{DBLP:conf/pakdd/WangCLZ13} presented a novel method named Multi-Manifold Ranking (MMR) to improve the efficiency of image retrieval. By using the content of the image files, Singh et al.~\cite{DBLP:Singh} proposed a novel framework for retrieving user's multimedia content like images from the user's profile. They used the Logical Item-set to mine on the image metadata consisting of the textual data associated with the images. For efficient medical image retrieval, Kitanovski et al.~\cite{DBLP:journals/mta/KitanovskiSDML17} developed an implemented system which performs image retrieval based on textual and image content. Zhao et al.~\cite{DBLP:conf/mm/ZhaoYYZ14} focused on affective image retrieval and investigate the performance of different features on different kinds of images in a multi-graph learning framework. Rasiwasia et al.~\cite{DBLP:conf/mm/RasiwasiaPCDLLV10} concentrated on the problem of joint modeling the text and image components of multimedia documents. They investigated two hypotheses and studied the task of cross-modal~\cite{DBLP:conf/sigir/WangLWZZ15} document retrieval which including the text that most closely match a query image, or retrieving the image that most closely match a query text. They used Gaussian Mixture Model to represent extracted visual features and applied a product quantization technique to provide fast response times over large image collections. Unfortunately, these studies aforementioned did not consider the images retrieval with spatial information such as longitude and latitude, as well as applying interactive searching. Thus they cannot solve the problem of interactive geo-tagged image search effectively.

Unlike the researches introduced above, there are some other researches in multimedia information retrieval~\cite{DBLP:journals/pr/WuWGL18,NNLS2018,DBLP:conf/pakdd/WangLZW14,TSMCS2018,YangJian14} area focused on image retrieval concerning geo-location to improve accuracy and efficiency. By using color layout descriptors and curvelet transform, Memon et al.~\cite{DBLP:journals/mta/MemonLMA17} proposed a method of Geo-location-based image retrieval to search for a set of color images match to the geo-location in the query image, but this approach did not apply geo-tags or location information of images. Liu et al.~\cite{DBLP:conf/mm/LiuHCSY12} presented a novel approach for discovering areas of interest (AoI) by analyzing both geo-tagged images and check-ins. They designed a novel joint authority analysis framework to rank AoI. To boost performance of the system, Yaegashi et al.~\cite{DBLP:conf/accv/YaegashiY10} utilized geotags consisting of values of latitude and longitude as additional information for visual recognition of consumer photos. To improving the efficiency of geo-tagged image retrieval, Kamahara et al.~\cite{DBLP:conf/mm/KamaharaNT12} attempted to retrieve images having certain similarities from a geo-tagged image dataset based on the concept of embedding geographical identification. To minimizing the disturbance by noisy tags, Xie et al.~\cite{DBLP:conf/icmcs/XieYH14} propose a hypergraph-based framework which integrates image content, user-generated tags and geo-location information into image ranking problem. These works exploited geo-location information of images to lifting accuracy and efficiency, however, they did not consider integrating interactive searching method into their systems to further enhance the accuracy of retrieval.

Content-based image retrieval~\cite{DBLP:conf/mm/WangLWZZ14,DBLP:conf/ijcai/WangZWLFP16,TII2018,DBLP:journals/pr/WuWLG18} (CBIR for short) using interactive search techniques is another issue focused by researchers in recent years. Bart Thomee and Michael S.Lew~\cite{DBLP:journals/ijmir/ThomeeL12} presented a review of interactive search in image retrieval aims to capture the wide spectrum of paradigms and methods in interactive search. Gosselin et al.~\cite{DBLP:journals/tip/GosselinC08} focused on statistical learning techniques for interactive image retrieval. They proposed the RETIN active learning strategy for interactive learning in content-based image retrieval context. Their strategy led to a fast and efficient active learning scheme to online retrieve query concepts from a database. However, these works are not suitable for the application scenario that the geographical proximity and visual content relevance have to be considered simultaneously.

\subsection{Spatial Keyword Queries and Interactive Search}
Spatial keyword queries is an important technique in the area of spatial database and information retrieval. It aims to search spatial objects combining geo-location information and keywords. Top-$k$ spatial keyword queries is one of the popular issues, which is studied by a large number of researchers recently. Ilyas I F et al.~\cite{DBLP:journals/csur/IlyasBS08} provided a classification for top-\emph{k} techniques based on several dimensions such as the adopted query model, data access, implementation level and supported ranking functions. Zhang et al. studied the problem of a fundamental spatial keyword queries called top-\emph{k} spatial keyword search~\cite{DBLP:conf/icde/ZhangZZL13,DBLP:journals/tkde/ZhangZZL16}. It retrieves the closest \emph{k} objects each of which contains all keywords in the query. They designed a novel index structure named inverted linear quadtree (IL-Quadtree) and proposed an efficient algorithm. Rocha-Junior J B et al.~\cite{DBLP:conf/edbt/Rocha-JuniorN12} considered the challenging problem of processing top-\emph{k} spatial keyword queries on road networks first time. Cong et al.~\cite{DBLP:journals/pvldb/CongJW09} proposed a new indexing framework for location-aware top-\emph{k} text retrieval. They used the inverted file for text retrieval and the R-tree for spatial proximity querying. Zhang et al.~\cite{DBLP:conf/sigir/ZhangCT14} processed distance-sensitive spatial keyword query as a top-\emph{k} aggregation query and designed a scalable and efficient solution based on the ubiquitous inverted list index. As users always interest in receiving up-to-date tweets such that their locations are close to a user specified location and their texts are interesting to user, Chen et al.~\cite{DBLP:conf/icde/ChenCCT15} studied the temporal spatial-keyword top-\emph{k} subscription (TaSK) query. They proposed a novel approach to efficiently process a large number of TaSK queries over a stream of geo-textual objects. Rocha-Junior J B et al.~\cite{DBLP:conf/ssd/RochaGJN11} proposed a novel index called Spatial Inverted Index ($S2I$) to improve the performance of top-\emph{k} spatial keyword queries. Their two algorithms named SKA and MKA support top-\emph{k} spatial keyword queries efficiently. Zhang et al.~\cite{DBLP:conf/edbt/ZhangTT13} proposed a scalable index named $I^\emph{3}$ for efficient top-\emph{k} spatial keyword search.

Another significant spatial keyword queries is called collective spatial keyword queries which aims to find a set of objects in the database such that it covers a set of given keywords collectively and has the smallest cost~\cite{DBLP:conf/sigmod/LongWWF13}. Cao et al.~\cite{DBLP:conf/sigmod/CaoCJO11} defined collective spatial keyword querying and studied two variants of this problem, then they proved that both of them are NP-complete. Long et al.~\cite{DBLP:conf/sigmod/LongWWF13} studied two types of the \emph{CoSKQ} problems, \emph{MaxSum-CoSKQ} and \emph{Dia-CoSKQ}. Zhang et al.~\cite{DBLP:conf/icde/ZhangCMTK09} presented a novel spatial keyword query problem called the \emph{m}-closest keywords (\emph{m}CK) query. They proposed a novel index named bR*-tree extended from the R*-tree to address this problem. The R*-tree designed by Beckmann et al.~\cite{DBLP:conf/sigmod/BeckmannKSS90}, which incorporates a combined optimization of area, margin and overlap of each enclosing rectangle in the directory. It outperforms the existing R-tree~\cite{DBLP:conf/sigmod/Guttman84} variants. As the exact algorithms for \emph{m}CK query are computationally expensive, Guo et al.~\cite{DBLP:conf/sigmod/GuoCC15} proved that answering \emph{m}CK query is NP-hard and designed two approximation algorithms called \emph{SKECa} and \emph{SKECa+}. On this basis, they proposed an exact algorithm utilizing \emph{SKECa+} to reduce the exhaustive search space significantly. Deng et al.~\cite{DBLP:journals/tkde/DengLLZ15} proposed a generic version of closest keywords search named best keyword cover which considers inter-objects distance as well as the keyword rating of objects. In order to improve the performance when the number of query keywords increases, they proposed a scalable algorithm called keyword-NNE which can significantly reduce the number of candidate keyword covers generated.

Unfortunately, all the researches aforementioned aim to query spatial objects with spatial information and keywords but they do not consider interactive search for images associated with geographical location, which is of equal importance to spatial keyword queries. These solutions for spatial keyword queries are really significant but they are not adequately suitable to the problem of interactive geo-tagged image query.

\section{Preliminary}
\label{preliminary}

In this section, we firstly review two basic approaches of image representation, namely invariant feature transform and bag-of-visual-words. Then we formally define the problem of interactive search for geo-tagged image and some concepts. Furthermore, we introduce the outline of our framework. Table ~\ref{tab:notation} summarizes the mathematical notations used throughout this paper to facilitate the discussion of our work.

\begin{table}
	\centering
    \small
	\begin{tabular}{|p{0.15\columnwidth}| p{0.73\columnwidth} |}
		\hline
		\textbf{Notation} & \textbf{Definition} \\ \hline\hline
		~$\mathcal{O}$                                   & A given database of geo-tagged image                                \\ \hline
        ~$o$                                             & A geo-tagged image object                                     \\ \hline
	  	~$q$                                             & A geo-tagged image query                                      \\ \hline
        ~$o.\varrho$                                     & The geo-location information descriptor of $o$                                 \\ \hline
		~$o.\psi$                                        & A visual content descriptor of $o$                              \\ \hline
		~$\bar{o}$                                       & A geo-tagged image skyline object             \\ \hline
        ~$\mathcal{X}$                                   & The longitude of a geo-location              \\ \hline
        ~$\mathcal{Y}$                                   & The latitude of a geo-location              \\ \hline
        ~$\emph{\textbf{p}}$                             & A preference vector                 \\ \hline
        ~$\hat{\emph{\textbf{p}}}$                       & An estimated preference vector       \\ \hline
        ~$k$                                             & The number of final results              \\ \hline
        ~$\vartheta$                                     & The maximum number of intermediate results displayed at each round  \\ \hline
        ~$\mathcal{F}_{score}(q,o)$                      & Score function measuring the relevance of $q$ and $o$   \\ \hline
        ~$\mathcal{F}_{prefer}(q,o,\emph{\textbf{p}})$   & Score function measuring the relevance of $q$ and $o$ with the preference \emph{\textbf{p}}  \\ \hline
        ~$\Theta (k,q,\mathcal{O})$                      & The $k$-superiors set of $\mathcal{O}$ by query $q$      \\ \hline
        ~$\mathcal{R}$                                   & The set of rounds during a interactive query      \\ \hline
        ~$|\mathcal{R}|$                                 & The total number of rounds during a interactive query      \\ \hline
        ~$r$                                             & The id of a round.                          \\ \hline
        ~$\mathcal{S}$                                   & The candidate set                                \\ \hline	
    	~$\mathcal{S}^r$                                 & The selected candidate to present to users at each found   \\\hline
        ~$\mathcal{C}$                                   & The set of constraint                                \\\hline
        ~$\mathcal{C}^i$                                 & The constraint obtained in round $i$                                \\\hline
	\end{tabular}
    \caption{The summary of notations} \label{tab:notation}	
\end{table}

\subsection{Review of Image Representation Techniques}
Image representation is one of the most significant problems in the area of multimedia retrieval, image recognition and computer vision. In this paper, we adopt scale invariant feature transform (SIFT for short) technique and bag-of-visual-words (BoVW for short) model for image representation, which are introduced in the following two subsections respectively.
\subsubsection{Scale Invariant Feature Transform}
Scale Invariant Feature Transform (SIFT) proposed by Lowe~\cite{DBLP:conf/iccv/Lowe99} is used as one of the basic methods for image representation, which aims to detects and describes local features in images~\cite{DBLP:journals/ijcv/Lowe04}. This approach transforms an image into a large collection of local feature vectors, each of which is invariant to image translation, scaling, and rotation, and partially invariant to illumination changes and affine or 3D projection. The features generated by SIFT share similar properties with neurons in inferior temporal cortex that are used for object recognition in primate vision~\cite{DBLP:conf/iccv/Lowe99}. This method is on the basis of a model of the behavior of complex cells in the cerebral cortex of mammalian vision. The procedure of image feature set generation includes four main phases~\cite{DBLP:journals/ijcv/Lowe04} which are introduced as follows:

\textbf{(1) Scale-space extrema detection.} In the first phase, the algorithm searches all scales and image location. By applying a difference-of-Gaussian (DoG) function $\mathcal{D}(x,y,\sigma)$ convolved with the image, which can be calculated from the difference of two nearby scales separated by a constant multiplicative factor $k$, it can recognize the potential interest points which are invariant to scale and orientation. $\mathcal{D}(x,y,\sigma)$  is defined as follows:

\begin{equation}\label{equ:igis DoG}
\mathcal{D}(x,y,\sigma)=\mathcal{L}(x,y,k\sigma)-\mathcal{L}(x,y,\sigma)
\end{equation}

where $\mathcal{L}(x,y,\sigma)$ is a scale-space function of an image $\mathcal{I}(x,y)$. It conducts the convolution of a variable-scale Gaussian core function $\mathcal{G}(x,y,\sigma)$ with an image $\mathcal{I}(x,y)$, as indicated in equation~\ref{equ:igis Scale} and~\ref{equ:igis gaussi} respectively:

\begin{equation}\label{equ:igis Scale}
\mathcal{L}(x,y,\sigma)=\mathcal{G}(x,y,\sigma) \ast \mathcal{I}(x,y)
\end{equation}

where $\ast$ is the convolution operation in $x$ and $y$, and

\begin{equation}\label{equ:igis gaussi}
\mathcal{G}(x,y,\sigma)=\frac{1}{2\pi \sigma^2}e^{-(x^2+y^2)/2}
\end{equation}

\textbf{(2) Keypoint localization.} The second phase aims to precisely localize the keypoints. A detailed model is fit to the nearby data to determine location, scale, and ratio of principal curvatures. at each candidate location. By measuring their stability, the keypoints are selected. In order to improve the stability of the keypoints, this phase uses the Tylar expansion of the scale-space function $\mathcal{L}(x,y,\sigma)$, shown as follows:

\begin{equation}\label{equ:igis tylar}
\mathcal{D}(\textbf{x})=\mathcal{D}+\frac{\partial{\mathcal{D}^T}}{\partial{\textbf{x}}}\textbf{x}+\frac{1}{2}\textbf{x}^T\frac{\partial^{2}\mathcal{D}}{\textbf{x}^2}\textbf{x}
\end{equation}

where $\mathcal{D}(\textbf{x})$ and its derivatives are evaluated at the sample point and vector $\textbf{x}=(x,y,\sigma)^T$ is the offset from the sample point. The extremum $\hat{\textbf{x}}$ can be calculated by taking the derivative of function $\mathcal{D}(\textbf{x})$ and with respect to $\textbf{x}$, when the value of function is set to 0, the extremum is

\begin{equation}\label{equ:igis extremum}
\hat{\textbf{x}}=-\frac{\partial^{2}\mathcal{D}^{-1}}{\partial{\textbf{x}^2}} \frac{\mathcal{D}}{\partial{\textbf{x}}}
\end{equation}

according to the extremum, the value of function~\ref{equ:igis tylar}, namely $\mathcal{D}(\hat{\textbf{x}})$ can be computing as

\begin{equation}\label{equ:igis valueoftylat}
\mathcal{D}(\hat{\textbf{x}})=\mathcal{D}+\frac{1}{2}\frac{\partial{D^T}}{\textbf{x}}\hat{\textbf{x}}
\end{equation}

The extremum points with a small $|\mathcal{D}(\hat{\textbf{x}})|$ are very unstable because they are sensitive to noise. When the value of $|\mathcal{D}(\hat{\textbf{x}})|$ is less than a empiric value, such as 0.03 (used in the work~\cite{DBLP:journals/ijcv/Lowe04}), the extremum point is discarded.

\textbf{(3) Orientation assignment.} In this phase, according to local image gradient directions, one or more orientations are designated to each keypoint. All future operations are performed on image data that has been transformed relative to the assigned orientation, scale, and location for each feature, thereby providing invariance to these transformations. For each image sample with the closest scale $\mathcal{L}(x,y)$, all computations are performed in a scale-invariant manner. At scale $\mathcal{L}(x,y)$ the gradient magnitude $m(x,y)$ and orientation $\theta(x,y)$ is computed as follows:

\begin{equation}\label{equ:igis magnitude}
m(x,y)=\sqrt{(\mathcal{L}(x+1,y)-\mathcal{L}(x-1,y))^2+(\mathcal{L}(x,y+1)-\mathcal{L}(x,y-1))^2}
\end{equation}

\begin{equation}\label{equ:igis orientation}
\theta(x,y)=tan^{-1}((\mathcal{L}(x,y+1)-\mathcal{L}(x,y-1))/(\mathcal{L}(x+1,y)-\mathcal{L}(x-1,y)))
\end{equation}

After computing the magnitude and orientation within a region around the keypoint, an orientation histogram can be formed, which has 36 bins covering the 360 degree range of orientations. The dominant orientation of the keypoint is designated by the highest peak in the histogram.

\textbf{(4) Keypoint descriptor.} In the last phase, this method measures local image gradients at the selected scale in the region around each keypoint. These local image gradients are transformed into a representation which allows for predominant levels of local shape distortion and illumination change. SIFT descriptor is a representation of gradients within the local image region. By partitioning the area around the keypoint and then forming the gradient histogram in this area, the keypoint descriptor can be generated.

\subsubsection{Bag-of-Visual-Words}
Recently, lots of image retrieval researches use the bag-of-visual-words (BoVWs for short) model to represent images, which is a extension from bag-of-words (BoW for short) technique using in text retrieval. BoVWs represents a image by a vector of visual vocabulary $\mathcal{W}$ which is constructed by vector quantization of feature descriptors~\cite{DBLP:conf/iccv/SivicZ03}. $k$-means is regularly used in this process to generate the visual vocabulary and the image database or a subset is used for training. The $k$-means cluster centers define visual words and the SIFT features in every image are then assigned to the nearest cluster center to give a visual word representation~\cite{DBLP:conf/bmvc/ChumPZ08}.

The size of a visual vocabulary $\mathcal{W}$ is denoted as $|\mathcal{W}|$, that means there are $|\mathcal{W}|$ visual words in it, denoted as $\{w_1,w_2,...,w_{|\mathcal{W}|}\}$. If a image is represented by a vector of length $|\mathcal{W}|$ applying BoVWs model, each element of the vector denotes the number of features in the image that are represented by given visual word. Let $\mathcal{A}_i$ is a visual words set and $\mathcal{A}_i \subseteq \mathcal{W}$, it is a weaker representation that does not store the number of features but only whether they are present or not. In others words, each element is binary.

Based on the conception above, two types of image similarity measures can be defined as follows:

\noindent\textbf{Set similarity.} Let $\mathcal{A}_1$ and $\mathcal{A}_2$ are two visual words set, and all the words are equally important. the similarity of $\mathcal{A}_1$ and $\mathcal{A}_2$ can be measured by:
\begin{equation}\label{equ:igis setsim}
Sim_s(\mathcal{A}_1, \mathcal{A}_2)=\frac{|\mathcal{A}_1 \cap \mathcal{A}_2|}{|\mathcal{A}_1 \cup \mathcal{A}_2|}
\end{equation}

\noindent\textbf{Weight set similarity.} Let $\mathcal{A}_1$ and $\mathcal{A}_2$ are two visual words set, $d_i \in \mathcal{D}$ be the weight of visual word $w_i$ and $d_i \geq 0$, the weight set similarity of $\mathcal{A}_1$ and $\mathcal{A}_2$ can be measured by:
\begin{equation}\label{equ:igis weightset}
Sim_w(\mathcal{A}_1, \mathcal{A}_2, \mathcal{D})=\frac{\sum_{w_i \in \mathcal{A}_1 \cap \mathcal{A}_2}^{}d_i}{\sum_{w_i \in \mathcal{A}_1 \cup \mathcal{A}_2}^{}d_i}
\end{equation}

It is obvious that set similarity is a particular case of weight set similarity if $\forall d_i \in \mathcal{D}, \exists d_i=1$. In this study we apply aforementioned similarity measures to interactive geo-tagged image search.

\subsection{Problem Definition}
\begin{definition}[\textbf{Geo-tagged Image Object}] \label{def:igis gio}
A geo-tagged image database can be defined as $\mathcal{O}=\{o_1, o_2,..., o_n\}$. Each geo-tagged image object $o \in \mathcal{O}$ is associated with a geo-location information descriptor $o.\varrho$ and a visual content descriptor $o.\psi$. A 2-dimensional geo-location with longitude $\mathcal{X}$ and latitude $\mathcal{Y}$ is represented by $o.\varrho = (\mathcal{X},\mathcal{Y})$, and $o.\psi = (w_1,w_2,...,w_n)$ is a visual words vector encoded into BoVWs model with millions of visual words.
\end{definition}

\begin{definition}[\textbf{Score Function}] \label{def:igis sf}
Let $q$ be a geo-tagged image query and $o$ be a geo-tagged image object. The score function $\mathcal{F}_{score}(q,o)$ is defined in detail below,
\begin{equation}
\mathcal{F}_{score}(q,o)=\lambda Dis(q.\varrho, o.\varrho)+(1-\lambda)Sim(q.\psi,o.\psi)
\end{equation}
where $Dis(q.\varrho, o.\varrho)$ is the distance function which normalizes the Euclidean distance between $q$ and $o$, $Sim(q.\psi,o.\psi)$ is the similarity function which measures the similarity or relevance between $q$ and $o$, and $\lambda \in [0,1]$ is a weighting parameter used to balance the geographical proximity and the image content similarity. Apparently, a high $\lambda$ indicates that users prefer the objects that are geographically closer to the query location while a low $\lambda$ explains users want to find the objects whose visual content is more relevant to visual content of the query. Before a query, users can be allowed to set their preferences quantified by $\lambda$ between the image content relevancy and the geo-location proximity.

As introduced in subsection 3.1, most content-based image retrieval technique based on the BoVWs model. An image, in general, is represented by a set of local visual features encoded into a vector containing visual words. Image represented by BoVWs model is a sparse high-dimension vector. Therefore we use set similarity algorithm to measure the similarity or relevance between $q$ and $o$, namely
\begin{equation}
Sim(q.\psi,o.\psi)=\frac{|q.\psi \cap o.\psi|}{|q.\psi \cup o.\psi|}
\end{equation}
\end{definition}
It is obvious that the score function can measure the similarity or relevance of two geo-tagged image objects. We apply it to top-$k$ geo-tagged image query which is defined as follows.

\begin{definition}[\textbf{Top-$k$ Geo-tagged Image Query}] \label{def:igis tkgiq}
Given a geo-tagged image database $\mathcal{O}$ and the number of results $k$, a top-$k$ geo-tagged image query $q_{_k}$ is define as $q_{_k}:(q_{_k}.\varrho, q_{_k}.\psi)$. Similar to Definition 1, $q_{_k}.\varrho = (\mathcal{X},\mathcal{Y})$ is a geo-location descriptor and $q_{_k}.\psi = (w_1,w_2,...,w_n)$ is a visual words vector encoded into BoVWs model, which contains $n$ visual words. It returns a set $\mathcal{S}$ of up to $k$ geo-tagged image objects from $\mathcal{O}$ such that they have the highest scores calculated by score function $\mathcal{F}_{score}(q_{_k}, o)$, i.e.,
\begin{equation}
\mathcal{S}=\{\mathcal{S} \subseteq \mathcal{O},|\mathcal{S}|=k \mid \forall o \in \mathcal{S}, o' \in \mathcal{O} \setminus \mathcal{S},\mathcal{F}_{score}(q_{_k}, o) \geq \mathcal{F}_{score}(q_{_k}, o')\}
\end{equation}
\end{definition}

The score function $\mathcal{F}_{score}(q,o)$ considers to balance the importance of geographical proximity and image content similarity by $\lambda$, but it neglects the importance or weight of each visual words that can reflect the preference of the users. Therefore, in order to solve the interactive geo-tagged image search problem, we propose a novel score function named preference function based on weight set similarity and $\mathcal{F}_{score}(q,o)$ to measure the similarity of objects with the user's preference.

\begin{definition}[\textbf{Preference Function}] \label{def:igis pf}
Let $q_{_k}:(q_{_k}.\varrho, q_{_k}.\psi)$ be a Top-$k$ geo-tagged image query specified by a user, $\emph{\textbf{p}}=(p_0,\emph{\textbf{p}}')$ be a preference vector which is used to represent user preference, in which $\emph{\textbf{p}}' = (p_1,p_2,...,p_i,...,p_n)$, $\forall i \in [1,n], 0 \leq p_i \leq 1$. For each geo-tagged image object $o \in \mathcal{O}$, the relevance between $q$ and $o$ can be measured by the following preference function,
\begin{equation}
\mathcal{F}_{prefer}(q_{_k}, o, \emph{\textbf{p}})= p_0 Dis(q_{_k}.\varrho, o.\varrho)+ Sim_w(q_{_k}.\psi,o.\psi,\emph{\textbf{p}}')
\end{equation}
\end{definition}

\begin{equation}
Sim_w(q_{_k}.\psi,o.\psi,\emph{\textbf{p}}')=\frac{\sum_{w_i \in q_{_k}.\psi \cap o.\psi}^{}p_i}{\sum_{w_i \in q_{_k}.\psi \cup o.\psi}^{}p_i}
\end{equation}

\begin{definition}[\textbf{Interactive Top-$k$ Geo-tagged Image Query}] \label{def:igis itkgiq}
Given a geo-tagged image database $\mathcal{O}$, an integer $k$ and an unknown user preference vector $\emph{\textbf{p}}$, the interactive top-$k$ geo-tagged image query will be processed in the fashion of several rounds. In each round, the system returns at most $\vartheta$ geo-tagged image objects to the user and asks her to choose the favourite one. When the interaction is terminated, the system will estimate the preference vector $\emph{\textbf{p}}$ of the user by analyzing her feedbacks and return a final results set $\mathcal{S}$ which contains $k$ objects by preference function $\mathcal{F}_{prefer}(q_{_k}, o, \emph{\textbf{p}})$, i.e.,
\begin{equation}
\mathcal{S}=\{\mathcal{S} \subseteq \mathcal{O},|\mathcal{S}|=k \mid \forall o \in \mathcal{S}, o' \in \mathcal{O} \setminus \mathcal{S},\mathcal{F}_{prefer}(q_{_k}, o) \geq \mathcal{F}_{prefer}(q_{_k}, o')\}
\end{equation}
\end{definition}

In the aforementioned definition, $\vartheta$ is defined as a constant to limit the number of image objects presented to the user, which is specified by human computer interface or psychological specialists to make users fell comfortable. It can be set by users according to their willing as well. In this paper, we propose a assumption that the preference of the user remains the same during the query process. In other words, at each round, the geo-tagged image object $o$ picked by the user must be the one whose the score calculating by $\mathcal{F}_{prefer}(q_{_k}, o, \emph{\textbf{p}})$ is the highest among the $\vartheta$ object displayed. We define this assumption formally as follows.

\newtheorem{assumption}{Assumption}
\begin{assumption}
Let $\textbf{N}^*$ be the set of positive integers, $\mathcal{R} \subseteq \textbf{N}^*$ be the set of rounds during a process of interactive top-$k$ geo-tagged image query, $r$ be the id of a round, $\mathcal{S}^r$ be the selected candidate to present to users at round $r$, $\hat{o}$ be an image object selected by the user. $\forall r \in [1,|\mathcal{R}|], \exists \mathcal{S}^r, \mathcal{F}_{prefer}(q_{_k}, \hat{o}, \emph{\emph{\textbf{p}}}) > \mathcal{F}_{prefer}(q_{_k}, o, \emph{\emph{\textbf{p}}}), \forall o \in \mathcal{S}^r \setminus \{\hat{o}\}$.
\end{assumption}

Our solution works well under this assumption and in this paper we do not consider the situation that the preference of the user changes when she browses several objects.

\subsection{Framework Overview}

\begin{figure}[thb]
\newskip\subfigtoppskip \subfigtopskip = -0.1cm
\centering
\includegraphics[width=0.55\linewidth]{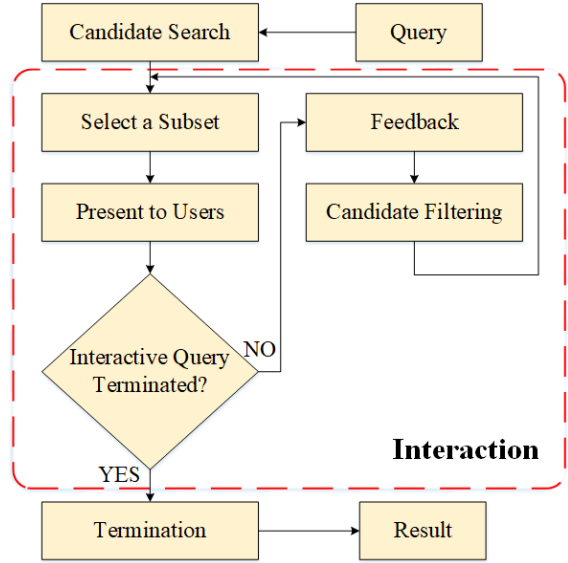}
\vspace{-1mm}
\caption{\small  The framework of the interactive geo-tagged image query }
\label{fig:fig2}
\end{figure}

We propose a framework named IGIQ$_k$ to solve the problem of interactive top-$k$ geo-tagged image query, which is demonstrated in Fig.~\ref{fig:fig2}. It consists of the following three stages:

\noindent\textbf{(1)Candidate Search.} In the candidate search stage, the system will search out a candidate set of $k$-superior geo-tagged image objects from the geo-tagged image database $\mathcal{O}$.

\noindent\textbf{(2)Interaction.} In the interaction stage, the system will interact with users in several rounds. At each round, the system will select a subset of candidates and display it to the user. When the user see these candidates, she can be allowed to choose the favourite object based on her personal preference. According to the selection of the user, the system will refine the candidate set and select another subset and present it to the user. In the meantime, once a termination condition is detected, the system will exist the interaction stage automatically.

\noindent\textbf{(3)Termination.} In the termination stage, the system estimates the preference vector $\emph{\textbf{p}}$ of the user and retrieval the top-$k$ image objects.

\section{Candidate Search Method}
\label{CandidateSearch}

In general, the volume of geo-tagged image database is considerably large, which contains a huge amount of image objects. When users input a query, it is not a efficient search strategy that the system traverses the whole database to find out a candidate set of objects at each round. In order to diminish the computing cost, we design a efficient method for candidate search by reducing the search space.

\subsection{Geo-tagged image object Superiority}
Before the start of the interaction, the system cannot acquire any information of user preference on geo-location and image content. Therefore, the candidate set generated in the candidate search stage should contain all the objects which could become a final result probably by a default preference.

Obviously, skyline operator proposed by Borzsonyi~\cite{DBLP:conf/icde/BorzsonyiKS01} is one of the significant notions in the database area. It aims to filters out a set of interesting points from a potentially large set of data points. A point is interesting if it is not dominated by any other point. A point dominates another point if it is as good or better in all dimensions and better in at least one dimension. Inspired by this notion, we propose a novel conception by extending the skyline to the geo-tagged image query area, namely geo-tagged image skyline, which servers as the basis of candidate search method. Geo-tagged image skyline is a set of objects in a geo-tagged image database that are superior to any other object in the both aspects of geo-location and relevance of visual content. Before defining geo-tagged image skyline formally, we introduce the notion of superiority relationship in the first instance, which is one of the key conceptions in this paper.

\begin{definition}[\textbf{Superiority Relationship}] \label{def:igis super}
Let $o_i$ and $o_j$ be two geo-tagged image objects in database $\mathcal{O}$. Given a geo-tagged image query $q$, the superiority relationship between $o_i$ and $o_j$ is defined as follow: under the query $q$, if $Dis(q.\varrho,o_i.\varrho) \leq Dis(q.\varrho,o_j.\varrho)$ and $Sim(q.\psi,o_i.\psi) \leq Sim(q.\psi,o_j.\psi)$, then $o_i$ is superior to $o_j$, denoted by $o_i \stackrel{_q}{\Rightarrow} o_j$. Otherwise, $o_i$ is not superior to $o_j$, denoted by $o_i \stackrel{_q}{\nRightarrow} o_j$.
\end{definition}

For a given geo-tagged image query $q$, we can measure the similarity or relevance between any object in database $\mathcal{O}$ and $q$. That is to say, the superiority relationship between these objects can be found with a query $q$. Therefore, Based on the definition of superiority and the notion of skyline, the geo-tagged image object skyline can be defined in the following form:

\begin{definition}[\textbf{Geo-tagged Image Skyline}] \label{def:igis gtis}
Given a geo-tagged image database $\mathcal{O}$ and a geo-tagged image query $q$, an object in $\mathcal{O}$ is a geo-tagged image skyline object, denoted as $\bar{o}$ if and only if $\forall o \in \mathcal{O}, o \stackrel{_q}{\nRightarrow} \bar{o}$.
\end{definition}

According to Definition \ref{def:igis gtis}, the geo-tagged image skyline $\bar{o}$ is the best result when we conduct a query $q$ on a database $\mathcal{O}$. In other words, it is the top-1 object query. For the top-$k$ query problem on a database $\mathcal{O}$, obviously the result set should contain $k$ objects which are superior to others in $\mathcal{O}$. Thus, we propose the notion named $k$-superiors based on geo-tagged image skyline.

\begin{definition}[\textbf{Geo-tagged Image $k$-Superior}] \label{def:igis gtiks}
Given a geo-tagged image database $\mathcal{O}$ and a geo-tagged image query $q$, an object is a geo-tagged image $k$-superior object in $\mathcal{O}$ denoted as $\tilde{o}$ if only if at most $k$ objects in $\mathcal{O}$ are superior to $\tilde{o}$. The set of $k$ superior objects is called $k$-superiors set of $\mathcal{O}$, denoted as $\Theta (k,q,\mathcal{O})$.
\end{definition}

\begin{lemma} \label{lemma:igis 1}
$\forall o \in \mathcal{O} \setminus \Theta (k,q,\mathcal{O})$, $o$ cannot belong to the top-$k$ results by preference function $\mathcal{F}_{prefer}(q_{_k}, o, \emph{\emph{\textbf{p}}})$ for any preference vector $\emph{\emph{\textbf{p}}}$.
\end{lemma}

Apparently, lemma \ref{lemma:igis 1} indicates that we can search the top-$k$ results according to the $k$-superiors set of $\mathcal{O}$. We prove lemma \ref{lemma:igis 1} based on definition \ref{def:igis gtiks} as follow:

\begin{proof}
According to Definition \ref{def:igis gtiks}, if $\forall o \in \mathcal{O} \setminus \Theta (k,q,\mathcal{O})$, then there are at least $k$ objects $\tilde{o}$ in database $\mathcal{O}$, which are superior to $o$. In other words, there are at least $k$ objects in $\mathcal{O}$ such that $\forall \emph{\textbf{p}}, \mathcal{F}_{prefer}(q_{_k}, \tilde{o}, \emph{\textbf{p}}) > \mathcal{F}_{prefer}(q_{_k}, o, \emph{\textbf{p}})$. Lemma~\ref{lemma:igis 1} is proved. $\blacksquare$
\end{proof}

\subsection{Candidate Search Algorithm}
In order to search out a candidate set without the guidance of preference, we design a novel index structure based on IR$^2$-Tree, namely Geo-tagged Image R-Tree (GIR-Tree) and propose a efficient search algorithm \textbf{G}eo-tagged \textbf{I}mage \textbf{SUPER}ior search algorithm (GI-SUPER for short) based on Definition~\ref{def:igis gtiks}.

\subsubsection{GIR-Tree Index}
\textbf{IR$^2$-Tree Index.} We propose a novel indexing structure GIR-Tree by extending IR$^2$-Tree to the geo-tagged image search area. IR$^2$-Tree is a significant indexing structure based on R-Tree~\cite{DBLP:conf/sigmod/Guttman84} in the area of spatial keyword search, which is proposed by Felipe et al.~\cite{DBLP:conf/icde/FelipeHR08} in 2008. It is a combination of an R-Tree and signature files. To be specific, each node of an IR$^2$-Tree contains both geo-location information and keywords. The geo-location information is represented in the form of a minimum bounding area (MBR for short) and the keywords are described as a signature. An IR$^2$-Tree is a height-balanced tree data structure, where each leaf node has entries of the form ($ObjPtr$, $A$, $Sig$). $ObjPtr$ and $A$ are defined as in the R-Tree while $Sig$ is the signature of the object referred by $ObjPtr$. A signature of a word is a fixed-length bit string created by a hashing function. The signature of a keyword set is the superimposition of all the signatures of these keywords. Fig 3 illustrates a simple IR$^2$-Tree.


\noindent\textbf{GIR-Tree Index.} It is obvious that each node of a IR$^2$-Tree contains two types of information: (1)geo-location information indicated in the form of $[latitude,longitude]$, and (2)signature of keywords. In order to describing a geo-tagged image object by a node, we design GIR-Tree in which a novel leaf node has the form ($ObjPtr$, $A$, $Sig$-$BoVW$). $ObjPtr$ and $A$ represent the geo-tagged image object pointer and geographical area respectively, and $Sig$-$BoVW$ is a signature generated from the BoVWs model of a geo-tagged image. A non-leaf node has entries of the form ($NodePtr$, $A$, $Sig$-$BoVW$), where $NodePtr$ refers to a node. As same as IR$^2$-Tree, GIR-Tree is a height-balanced tree as well. Fig.4 shows an example of GIR-Tree.


\subsubsection{Geo-tagged Image Superior Search Algorithm}

The pseudo-code of geo-tagged image superior (GI-SUPER) search algorithm is demonstrated in Algorithm 1. In the initial step, a $k$-superiors set $S$ is initialized as a empty set and $N_{root}$ is the root node of GIR-Tree of dataset. An empty min-heap $H$ is used to hold notes in GIR-Tree, which is to be visited. This min-heap is sorted descending by the value of sorting function $\mathcal{F}_{sort}(q,n)$. We define $n$ as a node of GIR-Tree with minimum bounding rectangle (MBR) and signature of BoVWs model denoted as $MBR(n)$ and $Sig\mbox{-}BoVW$ respectively. The detail of this funcion is shown as follow:

\begin{equation}
\mathcal{F}_{sort}(q,n)=Dis_{min}(q.\varrho, n.MBR(n))+Sim(q.\psi,n.Sig\mbox{-}BoVW)
\end{equation}

where the first term $Dis(q.\varrho, n.MBR(n))$ is the minimum distance between a query point and a MBR of a node, and the second term $Sim(q.\psi,n.Sig\mbox{-}BoVW)$ measures the similarity between the query image and the image contained in a node. It is easy to find that $\mathcal{F}_{sort}(q,n)$ is equivalent to $\mathcal{F}_{score}(q,n)$ when $\lambda = 0.5$ and distance is minimum.

In the iterative process, this algorithm firstly remove the top node $n$ from min-heap $H$ and then conducts the under-mentioned procedure according to the type of node until $H$ is empty:

\textbf{(1)} \textbf{If $n$ is a non-leaf note}. For each child note of $n$ the algorithm checks superiority relationship between $n$ and other $k$-superior objects to determine if there are at least $k$ objects which are superior to $n$. If $Dis(q.\varrho, o.\varrho) < Dis_{min}(q.\varrho, n.MBR(n))$ and $Sim(q.\psi,n.Sig\mbox{-}BoVW) < Sim(q.\psi,o.\psi)$ are satisfied at the same time, $o$ is superior to $n$. That means all the nodes in $n$ are farther away from $q$ than the object $o$ in the aspect of geo-location and the similarity between each node in $n$ and $q$ is smaller than the similarity between $o$ and $q$. If a child node passes the superiority relationship check, it is added to $H$. If not, it is abandoned.

\textbf{(2)} \textbf{If $n$ is a leaf node}. That means $n$ includes geo-tagged image objects. According to Definition~\ref{def:igis super}, the algorithm performs the superiority relationship check for each object by applying the geographical information and image visual content. If there are less than $k$ objects which are superior to an object $o_i$, it is added to $S$.

The iterative process is terminated if $H$ is empty. Obviously, $S$ is a $k$-superiors set of the dataset. That is, the objects selected and added to $S$ are the $k$-superior objects.

\begin{algorithm}
\begin{algorithmic}[1]
\footnotesize
\caption{\bf GI-SUPER Search Algorithm}
\label{alg:gt-super}

\INPUT  GIR-Tree index of a geo-tagged image dataset $T_{GIR}$, a query $q$, a constant $k$.
\OUTPUT A $k$-superiors set $S$.

\STATE Initializing: $S\leftarrow\emptyset$;
\STATE Initializing: An empty min-heap $H\leftarrow\emptyset$;
\STATE Initializing: $N_{root} \leftarrow$ root node of $T_{GIR}$;
\STATE $Add(H, N_{root})$;   //Add $N_{root}$ to $H$

\WHILE{$H \ne \emptyset$}
    \STATE $n \leftarrow $top node of $H$;
    \IF{$n$ is non-leaf node}
        \FOR{each child $n_i \in n$}
            \IF{less than $k$ objects in $S$ are superior to $n_i$}
                \STATE $Add(H,n_i)$;
            \ENDIF
        \ENDFOR
    \ELSE
        \FOR{each object $o_i \in n$}
            \IF{less than $k$ objects in $S$ are superior to $o_i$}
                \STATE $Add(S,o_i)$;
            \ENDIF
        \ENDFOR
    \ENDIF
\ENDWHILE
\RETURN $S$;
\end{algorithmic}
\end{algorithm}

\section{Interaction Method}
\label{Interaction}

After previous stage, a candidate set $\mathcal{S}$ can be return to users. Our approach then conduct a interaction with the users in rounds, which will select a subset from $\mathcal{S}$ and present them to the users. Our method will continuously refine preference of the users based on one favourite geo-tagged image object picking from $\mathcal{S}$ in current and all previous rounds. When the users want to stop it or the system automatically decides to exit, the interaction will be terminated. In order to search out $k$ objects which are preferred most by the users, the system will conduct interaction with sufficient rounds to test all pairs of candidates. Practically the scale of geo-tagged image database is very large and there are so many candidates in $\mathcal{S}$, that means the interaction requires lots of rounds. Therefore, we can propose the core problem in this stage is that how to select a candidate subset in each round so that the interaction can be terminated in a short time. In the following part, we propose the interaction algorithm at first. After that we propose the candidate subset selection method.

\subsection{Interaction Algorithm}
In this subsection we introduce the interaction algorithm shown by Algorithm~\ref{alg:interaction}. This algorithm aims to find out a set of constraints when the iteration is terminated based on a candidate set $\mathcal{S}$ and a query $q$. In the initialization part, the constraint set $\mathcal{C}$, the selected candidate set $\mathcal{S}^r$ to present to users at each found and a variable $i$ are initialized. Then the interaction is implemented by a loop. At the beginning of $i$-th round, the procedure $CandidateSelection(\mathcal{S},q)$ is conducted which aims to select a subset $\mathcal{S}^r$ from $\mathcal{S}$ according to the query $q$, which is a key process of this method. Then the subset $\mathcal{S}^r$ is shown to the user and she can choose the most favourite object $\hat{o}$ from $\mathcal{S}^r$ as the feedback. Subsequently the system generates a new constraint $\mathcal{C}^i$ based on the user's feedback by conducting procedure $ConstraintsGenerator(\hat{o})$. After the new constraint is added to the constraint set $\mathcal{C}$, the procedure $CandidateFilter(\mathcal{S},\mathcal{C})$ is conducted which improves $\mathcal{S}$ by analyzing the constraint set $\mathcal{C}$. At the end of this round, procedure $InteractionTerminate()$ is invoked, which aims to decide the iteration should be terminated or not. If it returns $true$, the interaction process is terminated immediately.

Next we present the candidate selection method and termination condition check method respectively, which are two core techniques of the interaction algorithm.

\begin{algorithm}
\begin{algorithmic}[1]
\footnotesize
\caption{\bf Interaction Algorithm}
\label{alg:interaction}

\INPUT  A geo-tagged image object candidate set $\mathcal{S}$, a query $q$.
\OUTPUT A constraint set $\mathcal{C}$.

\STATE Initializing: $\mathcal{C}\leftarrow\{\mathcal{C}^0\}$;
\STATE Initializing: $\mathcal{S}^r\leftarrow\emptyset$;
\STATE Initializing: $i\leftarrow1$;

\WHILE {$true$}
    \STATE $\mathcal{S}^r\leftarrow CandidateSelection(\mathcal{S},q)$;
    \STATE $CandidateShow(\mathcal{S}^r)$;  //Show the user the candidate objects
    \STATE $\hat{o}\leftarrow FavouriteSelection(\mathcal{S}^r)$  //the favourite object $\hat{o}$ is selected by the user
    \STATE $\mathcal{C}^i\leftarrow ConstraintsGenerator(\hat{o})$  //a new constraints is generated according to $\hat{o}$;
    \STATE Add $\mathcal{C}^i$ to $\mathcal{C}$
    \STATE $\mathcal{S}\leftarrow CandidateFilter(\mathcal{S},\mathcal{C})$
    \STATE $i$++;
    \IF {$InteractionTerminate()=true$}
        \STATE Break;
    \ENDIF
\ENDWHILE
\RETURN $\mathcal{C}$;
\end{algorithmic}
\end{algorithm}

\subsection{Candidate Selection Method}
The first step of the iteration presented above is candidate selection implemented which is a very important procedure. Different selection methods can generate different quality of preference approximation and search performance. In this section, we propose two type of selection methods, namely random selection method and densest subgraph method.

\subsubsection{Random Selection Method}
Random selection method is to generate the candidate subset $\mathcal{S}^r$ by selecting $\vartheta$ image objects randomly from $\mathcal{S}$, which is the most simple and straightforward method. However, this method has a shortcoming that it do not consider the superior relationship when the candidates are selected.

\subsubsection{Densest Subgraph Method}
For the purpose of maximizing the use of each round, the candidate set presented to the user $\mathcal{S}^r$ is the anticipated result when the expected number of constraints $|\mathcal{C}_{\mathcal{S}^r}|$ is maximized. According to our approach, the constraints can be generated when the user selects any object from $\mathcal{S}^r$ and feed it back to the system. Formally, $\mathcal{S}^r$ can be derived as follow,

\begin{equation}
\begin{split}
\mathcal{S}^r&=\argmax_{\mathcal{S}^r \subset \mathcal{S},|\mathcal{S}^r| \leq \vartheta}E[|\mathcal{C}_{\mathcal{S}^r}|] \\
&=\argmax_{\mathcal{S}^r \subset \mathcal{S},|\mathcal{S}^r| \leq \vartheta} \sum_{o_i \in \mathcal{S}^r}^{}P\{o_i\}*\mathcal{F}_{num}(o_i,\mathcal{S}^r)
\end{split}
\end{equation}
where $E[|\mathcal{C}_{\mathcal{S}^r}|]$ is the expected value of constraints number, $P\{o_i\}$ is the probability of $o_i$ selected by the user and function $\mathcal{F}_{num}(o_i,\mathcal{S}^r)$ is to compute the number of objects which are not superior to $o_i$ in $\mathcal{S}^r$. Let $\hat{\mathcal{S}^r}$ be the subset of $\mathcal{S}^r$, and $\forall o \in \hat{\mathcal{S}^r}, \nexists o' \in \mathcal{S}^r, o' \Rightarrow o$.That means the objects in $\mathcal{S}^r$ are not superior the objects in $\hat{\mathcal{S}^r}$. In this work, we suppose that the event of user choosing object from $\hat{\mathcal{S}^r}$ is an equally likely event. In formal, we define the probability as

\begin{equation}
P\{o_i\}=
\begin{cases}
\frac{1}{|\mathcal{C}_{\mathcal{S}^r}|},\quad \mbox{if} \quad o_i \in \mathcal{C}_{\mathcal{S}^r}\\
0,\quad \quad \mbox{otherwise}
\end{cases}
\end{equation}

Computing the optimal $\mathcal{S}^r$ is a quite complex task, thus we apply heuristic methods to improve the effectiveness and efficiency. Before introduce our method in detail, we propose a novel conception named no-superior graph.

\begin{definition}[\textbf{No-Superior Graph}] \label{def:igis no-superior graph}
Let $\mathcal{O}$ be a geo-tagged image database, no-superior graph of $\mathcal{O}$ is an non-directed graph, denoted by $\mathcal{G}_\mathcal{O} = (\mathcal{V},\mathcal{E})$, where $\mathcal{V}$ and $\mathcal{E}$ are the set of vertices and edges respectively. Each vertex $v_i \in \mathcal{V}$ represents an geo-tagged image object $o_i \in \mathcal{O}$ and each edge $e_{i,j} \in \mathcal{E}$ denotes that there are no superior relationship between vertices $v_i$ and $v_j$, namely $v_i \nRightarrow v_j$ and $v_j \nRightarrow v_i$.
\end{definition}

According to Definition~\ref{def:igis no-superior graph}, we can convert the problem of searching optimal $\mathcal{S}^r$ into the problem of searching the densest subgraph. The density of a graph can be defined as the ratio of the actual number of edges and max theoretical number of edges. We introduce it in a formal way below:

\begin{definition}[\textbf{Graph Density}] \label{def:igis density}
Let $\mathcal{G}=(\mathcal{V},\mathcal{E})$ be an non-directed graph, and the number of vertices and edges are denoted by $|\mathcal{V}|$ and $|\mathcal{E}|$. The density of $\mathcal{G}$ is defined as follows:
\begin{equation}
\xi(\mathcal{G})=\frac{2*|\mathcal{E}|}{|\mathcal{V}|*(|\mathcal{V}|-1)}
\end{equation}
\end{definition}

The study~\cite{DBLP:conf/waw/AndersenC09} indicate that finding the densest subgraph with at most $k$ vertices is a NP-Complete problem. But there are some approximate solutions that can solve this problem in polynomial time~\cite{DBLP:journals/siamcomp/GalloGT89,DBLP:A.V.Goldberg}. In our solution, only the vertices which are superior to others are considered when computing $\xi(\mathcal{G})$, rather than all the vertices in the graph. This is a profitable way to enhance computational efficiency. On the other hand, it is sufficient that searching the approximate densest subgraph with haphazard size because we do not have to obtain the optimal result.

Once the approximate densest subgraph has been searched out by our method, there are three possible situations must be considered: (1)$|\mathcal{S}^r| > \vartheta$; (2)$|\mathcal{S}^r|=\vartheta$ and (3)$|\mathcal{S}^r| < \vartheta$. Our algorithm will step into a loop and conduct different procedures for the different situations. Next we describe these three situations in detail:

(1)If $|\mathcal{S}^r| > \vartheta$, we will remove the objects which are superior to the most others in $\mathcal{S}^r$, until $|\mathcal{S}^r|=\vartheta$. Then the algorithm break the loop.

(2)If $|\mathcal{S}^r|=\vartheta$, we will attempt to remove the objects which are superior to the most others in $\mathcal{S}^r$ and test that $E[|\mathcal{C}_{\mathcal{S}^r}|]$ can be enhanced or not. If $E[|\mathcal{C}_{\mathcal{S}^r}|]$ can be enhanced, we will mark the removed objects as visited and continue the loop. Otherwise the algorithm break the loop.

(3)If $|\mathcal{S}^r| < \vartheta$, we will try to add one unvisited object from $\mathcal{S} \setminus \mathcal{S}^r$. Then we test that $E[|\mathcal{C}_{\mathcal{S}^r}|]$ can be enhanced or not. If $E[|\mathcal{C}_{\mathcal{S}^r}|]$ can be enhanced, the process will keep on conducting. Otherwise the algorithm break the loop.

The time complexity of this algorithm is linear because the times of loop is independent of $|\mathcal{S}|$.

\subsubsection{Candidate Filtering Method}
As shown in Algorithm~\ref{alg:interaction}, the procedure $CandidateFilter(\mathcal{S},\mathcal{C})$ will be conducted when a new constraint set $\mathcal{C}^i$ is generated and added into $\mathcal{C}$. This procedure aims to cut down the number of candidates which need to be considered in the next interaction. This method implements it by taking two actions on the no-superior graph as follows:

(1)For each adjacent vertex pair $(v_i,v_j)$, $v_i,v_j \in \mathcal{G}_\mathcal{O}$, if $v_i$ is superior or inferior to $v_j$ inferred from $\mathcal{C}^i$, the edge between $v_i$ and $v_j$, namely $e_{i,j}$ is removed.

(2)For each object $o_i \in \mathcal{S}$, if there are more than $k-1$ objects in $\mathcal{S}$ which are superior to $o_i$, then $o_i$ is removed.

\section{Termination Stage}
\label{Termination}

In the termination stage, our method decide that the interactive query processing should be terminated or not and then estimate the user preference vector $\emph{\textbf{p}}$. This stage contains two main procedure $InteractionTerminate()$ and $PreferenceEstimator()$, and we simply introduce these two procedure respectively in the following parts.

\subsection{Termination Procedure}
The main purpose of Termination procedure $InteractionTerminate()$ is to check whether the interaction query process should be terminated. Moreover, the users can decide that the interactive process is terminate or not as well according to their willing. With the interactive query processing going on, the uncertainty of user preference vector $\emph{\textbf{p}}$ can keep decreasing.

\subsection{Preference Estimation Procedure}
When the interactive query processing is terminated, a constraint set $\mathcal{C}$ is generated. The procedure $PreferenceEstimator()$ aims to find a estimator of $\emph{\textbf{p}}$, namely $\hat{\emph{\textbf{p}}}$ which subjects to $\mathcal{C}$. We design the estimation strategy enlightened by the support vector machine (SVM) techniques. The basic principle of SVM is to find a separation hyperplane that can assigns training examples to one category or the other by a apparent gap that is as wide as possible. Inspired by this, our method attempt to find a vector $\hat{\emph{\textbf{p}}}$ with the highest confidence to divide two categories: superior objects and inferior objects. This task can be converted into the process that minimizing the 2-norm of $\hat{\emph{\textbf{p}}}$, namely $||\hat{\emph{\textbf{p}}}||^2$ subject to all the constraints in $\mathcal{C}$. When we obtain the estimator of preference vector, the top-$k$ geo-tagged image query can be solved with $\hat{\emph{\textbf{p}}}$ to search the top-$k$ objects from the remaining candidates.

\section{Experimental Evaluation}
\label{Experiment}

In this section, we conduct a comprehensive experiments to evaluate the evaluate the effectiveness and efficiency of the proposed method in the paper.

\subsection{Experimental Settings}

\noindent\textbf{Workload.} A workload for this experiment includes 100 input queries. By default, the number of presented candidate $\varrho=8$, the number of query visual words $t = 100$, the number of final results $k=20$, and data number $N=60*10^4$. We use response time, precision, recall and F1-measure to evaluate the performance of the algorithms.

Experiments are run on a PC with Intel(R) CPU i7 7700K @4.20GHz and 16GB memory running Ubuntu 16.04 LTS Operation System. All algorithms in the experiments are implemented in Java.

\noindent\textbf{Dataset.} Our experiment aim to evaluate the performance of our solution against an image dataset which includes over one million geo-tagged images crawled from Flickr, a popular Web site for users to share and embed personal photographs. We used SIFT to extract all the local features of these geo-tagged images, then turn them into visual words represented as visual works vectors. The average number of local features is approximately 100.

\noindent\textbf{Baseline.} To our best knowledge, we are the first to study the problem of interactive geo-tagged image query. That means there are no existing method for this issue. We devise a baseline method without interaction, which applies inverted index and R-tree as the index structure based on the geographical information. The geo-tagged images are represented by visual words.

\noindent\textbf{Performance matric.} The objective evaluation of the proposed method is carried out based on response time, precision, recall and F1-measure. Precision measures the accuracy of search. Let $\alpha$ be the number of similar objects retrieved, $\beta$ be the total number of objects retrieved, $\gamma$ be the total number of similar objects in database. Precision is the ratio of retrieved objects that are relevant to the query.

\begin{displaymath}
Precision = \frac{\alpha}{\beta},\quad Recall=\frac{\alpha}{\gamma}
\end{displaymath}

\begin{displaymath}
F1-measure = \frac{2*Precision*Recall}{Precision+Recall}
\end{displaymath}

Recall measures the robustness of the retrieval. It is defined as the ratio of similar objects in the database that are searched in response to a query. And F1-measure is a comprehensive measure combining precision and recall. Higher F1-measure indicates that the system is more robust.

\subsection{Experimental Results}

\subsubsection{Evaluation on Interaction Stage}


\noindent\textbf{Evaluation on different number of results $k$.} We evaluate the two candidate selection methods proposed in this paper, namely random selection method and densest graph method on different number of results $k$. We increase $k$ from 5 to 100. We investigate the response time, precision, Recall and F1-mesuare in Fig, where other parameters are set to default values. Fig.(a) illustrates that the response time per round of random selection fluctuate between 5 and 10 but the time cost of densest graph increases gradually with the rising of $k$. When $k>20$, the growth is faster. We can see from the Fig.(b) that the precision of these two methods is going down step by step but the performance of densest graph method is much better than random selection. At $k=5$, the precision of densest graph is nearly 95\%. Fig.(c) depicts the recall tendency. Apparently, with the increasing of $k$, the recall of them are progressively climbing. Likewise, the recall of densest graph is higher than the other, which is close to 99\% when $k=100$. F1-measure shown in Fig.(d) describes the comprehensive performance of these two methods. It is easy to find that the F1-measure of densest graph is higher. In other words, the performance of densest graph is superior than random selection.


\noindent\textbf{Evaluation on different number of query visual keywords.} In order to evaluate the performance of random selection method and densest graph method under queries with different number of query visual keywords, we increase the number of visual keywords from 10 to 200. The response time per round of these methods are shown in Fig.(a). It is clearly that the efficiency of random selection is not effected by the increasing of query visual keywords. However, the time cost of densest graph method grows gradually. In Fig.(b) we can find that the precision of densest graph is higher than random selection, and it gently descends with the number of query keywords increasing. On the other hand, the downtrend of random selection is more evident. Fig.(c) indicates that both of the recall are climbing and performance of densest graph is better than random selection. As shown in Fig.(d), the overall performance of densest graph method is higher than random selection, which fluctuates in interval of 89\% to 93\%. On the other side, the f1-measure of random selection drops from 64\% to 43\%.


\noindent\textbf{Evaluation on different number of presented candidates $\vartheta$.}  We increase the number of presented candidates $\vartheta$ from 2 to 10. Fig.(a) demonstrates the time cost of random selection and densest graph. With the rising of $\vartheta$, the response time of densest graph decreases until $\vartheta$ increases to 8. After that it ascends to 35. Obviously, the response time of random selection is lower than the other, which fluctuates moderately. The precision of these two methods are shown in Fig.(b). It is not hard to see that both of them grow gradually with the increment of $\vartheta$. Likewise, the precision of random selection is lower than densest graph method. In Fig.(c), the recall of random selection and densest graph are slackly falling, and the f1-measure of densest graph is larger than random selection, indicated in Fig.(d). Overall, densest graph method has a better performance.

\subsubsection{Evaluation on Entire Query Processing}


\noindent\textbf{Evaluation on different number of results $k$.} We evaluate the performance of our approach named IGIQ$_k$ and the baseline with the increment of $k$. IGIQ$_k$ is the entire solution including three stage processing above-mentioned. In this experiment, we choose GI-SUPER Search algorithm in the candidate search stage and densest graph method in the interaction stage as the key techniques. Fig.(a) indicates that the precision of IGIQ$_k$ is higher than baseline. The former fluctuates between 96\% and 87\% and have a decreasing trend along with the rising of $k$.

\noindent\textbf{Evaluation on different number of query visual keywords.} To evaluate the precision of these two approaches, we increase the number of query visual keywords from 10 to 200. As shown in Fig.(b), the precision of IGIQ$_k$ rises and falls mildly between 89\% to 95\% and baseline goes down gradually. There is no doubt our method has a better performance in the aspect of precision than baseline.

\section{Conclusion}
\label{Conclusion}

In this paper, we study a novel research problem named interactive top-$k$ geo-tagged image query. Given a query image with geographical information, a interactive top-$k$ geo-tagged image query can learn the user's preferences based on the feedback and finally returns the $k$ best geo-tagged image objects ranked in terms of visual content similarity or relevance to the query objects and the spatial distance to the query location. Firstly we define this query in formal and then propose the framework of our approach, which includes three main stages. In order to improve the efficiency of searching, we present a novel index structure called GIR-Tree and a efficient algorithm named GI-SUPER algorithm to generate a small candidate set. To implement the interaction method that can learn user preferences precisely and efficiently, we propose the interaction framework and introduce two candidate selection methods. Besides, we discuss how to estimate the user preferences from the constraints generated from interaction and introduce the termination procedure briefly. The experimental results illustrate that our solution have a really high performance.

\textbf{Acknowledgments:} This work was supported in part by the National Natural Science Foundation of China
(61472450, 61702560), the Key Research Program of Hunan Province(2016JC2018), and project (2018JJ3691) of Science and Technology Plan of Hunan Province.



\bibliographystyle{spmpsci}      

\bibliography{ref}

\end{document}